\documentclass{amsart}
\usepackage{mathrsfs}
\usepackage[dvipdfmx]{graphicx}
\usepackage{cite}
\title[First--Quantized Theory of
Expanding Universe]{First--Quantized Theory of Expanding Universe
from Field Quantization in Mini--Superspace}
\author{Daisuke Ida}
\author{Miyuki Saito}
\date{Tuesday 12 August, 2014}
\begin{document}
\maketitle

\vspace{-20pt}
\begin{center}
{\small\it Department of Physics, Gakushuin University, Tokyo 171-8588}
\end{center}

\begin{abstract}
We propose an improved variant of the third--quantization scheme,
 for the 
spatially homogeneous and isotropic cosmological models
in Einstein gravity coupled with a neutral massless scalar field.
Our strategy is to specify a semi--Riemannian structure on the 
mini--superspace and to consider the quantum Klein--Gordon
field on the mini--superspace.
Then, the Hilbert space of this quantum system becomes inseparable,
which causes the creation of infinite number of universes.
To overcome this issue, we introduce a vector bundle structure on the
Hilbert space and the connection of the vector bundle.
Then, we can define a consistent unitary time evolution of the quantum universe
in terms of the connection field on the vector bundle.
By doing this, we are able to treat the quantum dynamics of 
a single--universe state.
We also find an appropriate observable set constituting the CCR--algebra,
and obtain the Schr\"odinger equation for the wave function of the 
single--universe state.
We show that the present quantum theory correctly reproduces the classical 
solution to the Einstein equation.
\end{abstract}

\section{Introduction}
One fundamental problem in quantum cosmology is the issue on the observable set
and the quantum state of the space-time.
The purpose of this paper is to construct an appropriate set of 
observables in 
the mini--superspace model of quantum cosmology and quantum states of the universe
with clear classical--quantum correspondence. 

While the perturbative quantization of the general relativity confronts the
difficulty due to the non--renormalizability of the theory, alternative approaches 
such as string theory, loop gravity, etc. 
have been widely discussed for a long time.
Needless to say, an important application of the quantum gravity would
be in the quantum cosmology, 
which would give quantum descriptions of the expanding universe at early times 
and the big-bang singularity.
Although we don't have a fully reliable theory of quantum gravity,
it will be worthwhile to discuss a quantum theory of the expanding universe,
such as the mini--superspace model, 
defined only on the finite dimensional reduced phase space of the gravitational field.

In the Hamiltonian formulation of the general relativity,
we are led,
according to the Dirac's prescription~\cite{Dir64} for singular Lagrangian systems,
 to the constrained Hamiltonian system, in which the 
total Hamiltonian of the gravitational field 
\begin{align*}
  H_T=\int_{\Sigma_t} d^3 x(N \Phi_0+\sum_{i=1}^{3}N^i\Phi_i)
\end{align*}
consists of 
a linear combination of first--class constraint functions
$\Phi_\mu\approx 0$ $(\mu=0,1,2,3)$
~\cite{Dir58,ADM62}.

The configuration space of the gravitational field
is parametrized by the Riemannian 3-metric $h_{ij}$ 
on the Cauchy surface $\Sigma_t$, and the canonical momentum is
given by 
\begin{align*}
  \pi^{ij}=(16\pi G)^{-1}\sqrt{h}(K^{ij}-Kh^{ij}),
\end{align*}
where $K_{ij}=(1/2)\mbox{\pounds}_n h_{ij}$
($n$ is the future pointing unit normal vector field on $\Sigma_t$)
 is the extrinsic curvature of $\Sigma_t$.
In terms of these canonical variables, the constraints  are written as
\begin{align*}
\Phi_0&=G_{ij,kl}\pi^{ij}\pi^{kl}-\dfrac{\sqrt{h}\mathscr{R}}{16\pi G}\approx 0,\\
\Phi_i&=-2 D_k \pi^{k}_i\approx 0,
\end{align*}
where $\mathscr{R}$ and $D_k$ are
the scalar curvature and  the covariant derivative (for tensor densities) on $(\Sigma_t,h_{ij})$, respectively, 
and  
\begin{align*}
  G_{ij,kl}:=\dfrac{8\pi G}{\sqrt{h}}\left(
h_{ik}h_{jl}
+h_{il}h_{jk}
-h_{ij}h_{kl}
\right)
\end{align*}
determines the semi--Riemannian structure of the 
configuration space of the 3-metrics.

According to the Dirac's algorithm of quantization,
each first--class constraint function $\Phi_\mu$ 
is replaced with a corresponding
linear operator $\widehat\Phi_\mu$ in a Hilbert space $\mathscr{H}$.
Then, 
the observed physical state is required to 
be a ray in
the subspace
$\mathscr{H}_{\rm phys}=\{|\psi\rangle\in \mathscr{H};\widehat\Phi_\mu|\psi\rangle=0\}$ 
determined by these constraint operators.
These requirements: $\widehat \Phi_\mu|\psi\rangle=0$,
for the physical state, which become functional differential equations
in the Schr\"odinger representation, are well known as the Wheeler--DeWitt equations~\cite{DeW67,Whe68}.

This formulation of quantum gravity is purely kinematical, in that the Wheeler--DeWitt
equations only make a restriction  on the physical Hilbert space, but does not define the
dynamics of the wave function, since the Hamiltonian itself becomes a constraint operator:
$\widehat H_T|\psi\rangle=0$, which is called the problem of time in the literature~\cite{Rov91,Kuc92,Ish93}. 
Furthermore, we don't have a fully satisfactory  interpretation of the quantum state
subject to the Wheeler--DeWitt equations.

In order to approach such conceptual issues, it will be worthwhile,
 as a first step, to discuss the
mini--superspace model of the quantum cosmology.
According to the concept of the geometrodynamics by Wheeler~\cite{Whe68}, 
the solution to the
Einstein equation can be viewed as the motion of a test particle in the superspace~\cite{DeW70,Mis72}, which
is the space of the 3--geometries on the Cauchy surface $\Sigma_t$, 
more precisely, 
it is the configuration space of 3-metrics above modulo the proper subgroup of the
 self--diffeomorphism group of $\Sigma_t$.
Since this viewpoint is very useful in our formulation of the quantum theory of cosmology, 
let us respect  it in the mini--superspace model in what follows. 

The organization of the paper is as follows.
In section~\ref{sec:pre}, we briefly review the Hamiltonian formulation
of the classical Einstein gravity  applied to the
Friedmann--Robertson--Walker (FRW) space--time,
where we introduce the semi--Riemannian metric $g_C$ on the 
mini--superspace.
In section~\ref{sec:KG}, we consider the quantum Klein--Gordon field
in the mini--superspace with respect to the semi--Riemannian structure 
given by $g_C$, where we are faced with the issue regarding the 
inseparability of the Hilbert space in the cases of the spatially
non--flat universe.
In section~\ref{subsec:hyp} and ~\ref{subsec:ell}, 
we introduce the vector bundle structure
in the inseparable Hilbert space to define the consistent quantum theory
with the unitary time evolution described by the global section of the
vector bundle, where we obtain the first--quantized theory of the
expanding universe on the space of 1--particle states of the quantum
Klein--Gordon field.
In section~\ref{sec:c-q}, we show that our quantum theory correctly
reproduces
the classical solution to the Einstein equation.
In section~\ref{sec:final}, we give  concluding remarks.

\section{Preliminaries: Mini--Superspace Model}\label{sec:pre}
In the mini--superspace model of the isotropic and homogeneous universe,
we consider the class of space--times represented by the FRW metric
\begin{align*}
  g=-N(t)^2dt^2+a(t)^2\gamma_K,
\end{align*}
where $\gamma_K$ ($K=0,\pm 1$) denotes the Riemannian metric of the 3--space of the constant 
sectional curvature $K$.
Let us consider the neutral massless scalar field 
minimally coupled with the Einstein--Hilbert action, which is provided by the classical action
\begin{align*}
  S[g,X]=\dfrac{1}{16\pi G}\int d^4 x\sqrt{-g}\left(R-\dfrac{1}{2}X_{,\mu}X^{,\mu}\right).
\end{align*}

In the standard procedure, we are led to the total Hamiltonian
\begin{align}
  H_T&=N\Phi_0,\label{eq:classicalhamiltonian}\\
\Phi_0&=
\dfrac{8\pi G}{v}\left(-\dfrac{p_a^2}{12a}+\dfrac{p_X^2}{a^3}\right)
-\dfrac{3Kva}{8\pi G}\nonumber, 
\end{align}
where we consider the fixed spatial coordinate volume and
set $v=\int d^3 x\sqrt{\gamma_K}$.
The configuration space $\mathscr{M}$,
which is called the mini--superspace,
is parametrized by $q^m=(a,X)$, and their conjugate momenta are denoted as
$p_m=(p_a,p_X)$.
The momentum constraints $\Phi_i\approx 0$ are automatically satisfied
in this parametrization.
As the Hamiltonian can be written in the form:
\begin{align*}
  H_T&=N\left[(g_S)^{mn}p_mp_n+u\right],\\
g_S&=\dfrac{v}{8\pi G}(-12 a da^2+a^{3}dX^2),\\
u&=-\dfrac{3Kva}{8\pi G},
\end{align*}
the dynamics of the universe is equivalent to that of a
point particle in the curved space--time with the metric $g_S$
and the potential function $u$.
In fact, the equation of motion is given by
\begin{align*}
  \dfrac{d^2 q^k}{dt^2}+\Gamma[g_S]^k_{mn}\dfrac{d q^m}{dt}\dfrac{dq^n}{dt}&
\approx
N^{-1}\dfrac{dN}{dt}\dfrac{dq^k}{dt}
-2N^2(g_S)^{kl}\dfrac{du}{dq^l},\\
\Phi_0=\dfrac{1}{4N^2}(g_S)_{mn}\dfrac{dq^m}{dt}\dfrac{dq^n}{dt}+u
&\approx 0.
\end{align*}

The present system has an invariance under the
coupled
conformal transformation of the Lorentzian metric $g_S$ 
and 
the pointwise scaling of the potential: 
$(g_S,u)\mapsto (fg_S,f^{-1}u)$ for  $f(q^m)\ne 0$.
Using this invariance, the system  can always be reduced to
that of the  geodesic particle in $\mathscr{M}$ but with
a conformally related metric~\cite{DeW70,Mis72}.

In fact,  assuming $u\ne 0$,
under the conformal transformation
\begin{align}
(g_S  )_{mn}=\dfrac{C^2}{u}(g_C)_{mn}\label{eq:gc}
\end{align}
and the reparametrization of the time function
\begin{align*}
  t(s)=\int^s ds\dfrac{C}{2Nu},
\end{align*}
the equations of motion become
\begin{align}
  \dfrac{d^2 q^k}{ds^2}+\Gamma[g_C]^k_{mn}\dfrac{d q^m}{ds}\dfrac{dq^n}{ds}&
\approx 0,\label{eq:geodesic}\\
\Phi_0=u\left((g_C)_{mn}\dfrac{dq^m}{ds}\dfrac{dq^n}{ds}+1\right)
&\approx 0,\label{eq:unitvector}
\end{align}
where $C$ is a constant with the mass dimension 1.
The Eq.~(\ref{eq:geodesic}) is the geodesic equation with respect
to the semi--Riemannian metric $g_C$,
and the Hamiltonian constraint~(\ref{eq:unitvector}) requires that
the new parameter $s$ is the proper time.
Hence, the equivalent system is given by the
Hamiltonian
of the geodesic particle in $(\mathscr{M},g_C)$,
\begin{align*}
  H'_T=\lambda \left((g_C)^{mn}p_mp_n+1\right),
\end{align*}
where $\lambda$ is the Lagrange multiplier.

On the other hand,  when $u=0$, 
we set $g_C=g_S$.
Then, 
the Hamiltonian is simply given by
\begin{align*}
  H'_T=N (g_C)^{mn}p_mp_n.
\end{align*}
This describes the null geodesic particle in $(\mathscr{M},g_C)$.

As a quantized system corresponds to the classical system defined 
by Eq.~(\ref{eq:classicalhamiltonian}),
we consider the quantum fields in the mini--superspace $(\mathscr{M},g_C)$
as a semi--Riemannian manifold.

\section{Quantum Klein--Gordon field in mini--superspace}\label{sec:KG}

Here, we consider the Klein--Gordon field in $(\mathscr{M},g_C)$
and give a consistent quantized theory.
Let us see $K=0,\pm 1$ cases separately, since they are technically 
different.

\subsection{Case of flat universe $(K=0)$:}\label{subsec:flat}
The spatially flat case turns out to be the most simple, so that it is appropriate
to explain this case first.
In this case, the dynamics of the universe is equivalent to
that of a null geodesic particle in $(\mathscr{M},g_C)$,
where the metric is given by
\begin{align*}
  g_C=\dfrac{v}{8\pi GC}(-12 a da^2+a^{3}dX^2).
\end{align*}
We construct the localized 1--particle states of the massless quantum Klein--Gordon field 
to describe a quantum mechanical counterpart of this  classical light--like particle.

Firstly, we introduce the new time function $T$ by
$a=a_0 e^{\beta T}$.
Then, the metric on $\mathscr{M}$ becomes
the conformally flat form
\begin{align*}
g_C=A^2 e^{3\beta T}(-dT^2+dX^2),
\end{align*}
where we set
\begin{align*}
  \beta=(12)^{-1/2},~~~A^2=\dfrac{va_0^3}{8\pi GC}.
\end{align*}

The classical action of the massless Klein--Gordon field 
in $(\mathscr{M},g_C)$
becomes that in the flat space:
\begin{align*}
  S[\phi]=\dfrac{1}{2}\int dT  dX \left[
(\partial_T\phi)^2-(\partial_X\phi)^2
\right].
\end{align*}
A solution to the Klein--Gordon equation can be decomposed into 
a linear combination of the mode functions $\{f(p;T,X),f^*(p;T,X)\}$
$(p\in\boldsymbol{R})$,
where 
\begin{align*}
  f(p;T,X)=(4\pi |p|)^{-1/2}e^{-i|p|T}e^{ipX}.
\end{align*}

The quantum Klein--Gordon field is expanded into the form
\begin{align*}
\phi (T,X)=\int_{-\infty}^\infty dp~ [a (p) f(p;T,X)+a^*(p) f^*(p;T,X)],
\end{align*}
and the conjugate momentum operator is written as
$ \pi(T,X)=\partial_T\phi(T,X)$.
The canonical commutation relations (CCRs) 
\begin{align*}
[\phi(T,X),\pi(T,X')]_-&
=i\delta(X-X'){\boldsymbol 1},\\
[\phi(T,X),\phi(T,X')]_-&=0,~~
[\pi(T,X),\pi(T,X')]_-=0,
\end{align*}
are equivalent to
\begin{align*}
  [a(p),a^*(p')]_-&=\delta(p-p'){\boldsymbol 1},~~
[a(p),a(p')]_-=0,~~~
[a^*(p),a^*(p')]_-=0.
\end{align*}
In terms of these, the quantum Hamiltonian operator is given by
\begin{align*}
  H=\int_{-\infty}^\infty dp~ |p|a^*(p)a(p).
\end{align*}

As usual, the vacuum state $|\Omega\rangle$ is defined by the requirement:
\begin{align*}
  a(p)|\Omega\rangle =0, ~~~\mbox{for all $p\in\boldsymbol{R}$},
\end{align*}
and the Fock space is constructed in the standard procedure.
Let $\mathscr{F}_\Omega^{(1)}$ denote the Hilbert space of 
1--particle states, which we are mainly concerned with.

Define, for each $p\in\boldsymbol{R}$, the momentum operator
\begin{align*}
  P:=\int_{-\infty}^\infty dp~pa^*(p) a(p),
\end{align*}
which is a Hermitian operator on $\mathscr{F}^{(1)}_\Omega$.
The eigenvectors of $P$ are given by
\begin{align*}
  |p\rangle:= a^*(p)|\Omega\rangle,~~~\mbox{for $p\in\boldsymbol{R}$},
\end{align*}
which satisfy
\begin{align*}
  P|p\rangle=p|p\rangle.
\end{align*}
These satisfy the orthogonality condition
\begin{align*}
  \langle p|p'\rangle=\delta(p-p')
\end{align*}
and the completeness condition
\begin{align*}
  \int_{-\infty}^\infty dp~|p\rangle\langle p|=\boldsymbol{1},
\end{align*}
where $\boldsymbol{1}$ denotes the identity operator on $\mathscr{F}^{(1)}_\Omega$.

The position operator is defined by
\begin{align*}
  Q:=i\int_{-\infty}^\infty dp~a^*(p)\partial_pa(p),
\end{align*}
which is a Hermitian operator on $\mathscr{F}^{(1)}_\Omega$.
The eigenvector of $Q$ is formally written as a  state 
\begin{align*}
  |X\rangle_Q:=\int_{-\infty}^\infty dp ~e^{-ipX}|p\rangle, ~~~
\mbox{for $X\in\boldsymbol{R}$},
\end{align*}
and it satisfies
\begin{align*}
  Q|X\rangle_Q=X|X\rangle_Q.
\end{align*}
The vector $|X\rangle_Q$ is 
the localized state considered by Newton and Wigner~\cite{NW49} 
long time ago.

The operators $P$ and $Q$ constitute the CCR--algebra
\begin{align*}
  [Q,P]_-=i\boldsymbol{1}.
\end{align*}
Thus, we obtain the canonical observable set $(P,Q)$ and its
Schr\"odinger representation on $\mathscr{F}^{(1)}_\Omega$. 

We regard the quantum state of the universe
as being described by a 1--particle state 
of the Fock space.
Since the Hamiltonian of the Klein--Gordon field 
acts on $\mathscr{F}_\Omega^{(1)}$,
any 1--particle state remains in $\mathscr{F}_\Omega^{(1)}$ 
in the course of the unitary time evolution.
In fact, the Hamiltonian  restricted
on $\mathscr{F}_\Omega^{(1)}$
can be written as 
\begin{align*}
  H^{(1)}= |P|.
\end{align*}
The expansion of the 1--particle state 
at the time $T$
\begin{align*}
|\psi(T)\rangle =\int_{-\infty}^\infty dp~ \psi(p,T)|p\rangle  
\end{align*}
defines the wave function $\psi(p,T)$ of the universe in the 
momentum representation.
Then, the time evolution of the
1--particle state
is determined by the Schr\"odinger equation
\begin{align*}
  i\partial_T \psi(p,X)=\langle p|H^{(1)} |\psi(T)\rangle
=|p|\psi(p,T),
\end{align*}
according to the unitary time evolution of the quantum Klein--Gordon field.
In this way, we obtain a quantum Hamiltonian for the wave function of
the universe without operator ordering ambiguity~\cite{HP86},
which clearly describes the massless particles  in $(\mathscr{M},g_C)$,
besides the well--defined canonical observable set $(P,Q)$.

\subsection{Case of hyperbolic universe $(K=-1)$:}\label{subsec:hyp}
In this case, 
we are led to the massive field in the expanding chart of the Milne space. 
Then, according to the standard approach,
the quantum field theory in this background geometry describes the continuous pair creations
of scalar particles, which is hard to interpret in the context of the quantum
cosmology. 
This occurs due to the inseparability of the Hilbert space involved.
Nevertheless, we show that it is possible to construct a consistent
quantum theory which describes the dynamics of a 1--particle state
with a correct classical interpretation.

The semi--Riemannian metric on $\mathscr{M}$ in this case is given by [See Eq.~(\ref{eq:gc})]
\begin{align*}
  g_C=
\dfrac{3v^2}{(8\pi G)^2C^2}(-12  a^2da^2+a^{4}dX^2).
\end{align*}
Introducing the new time function $T$ by $a=a_0e^{\beta T}$,
the metric is written as
\begin{align*}
  g_C=A^2e^{4\beta T}(- dT^2+dX^2),
\end{align*}
where we set
\begin{align*}
  \beta=(12)^{-1/2},~~~A=\biggl|\dfrac{\sqrt{3}va_0^2}{8\pi GC}\biggr|.
\end{align*}
Now, we start with the action of the massive Klein--Gordon field
\begin{align*}
  S[\phi]=\dfrac{1}{2}\int dT dX [(\partial_T\phi)^2-(\partial_X\phi)^2-A^2m^2e^{4\beta T}\phi^2],
\end{align*}
of the mass $m$.
The complete set of the solutions to the Klein--Gordon equation is given by
$\{f(p;T,X),f^*(p;T,X)\}$ $(p\in\boldsymbol{R})$, where
\begin{align*}
  f(p;T,X)&:=(4\pi |p|)^{-1/2}F(|p|;T)e^{ip X},\\
F(|p|;T)&:=\Gamma(1-i(2\beta)^{-1}|p|)\left(\dfrac{Am}{4\beta}\right)^{i(2\beta)^{-1}|p|}
J_{-i(2\beta)^{-1}|p|}\left(\dfrac{Ame^{2\beta T}}{2\beta}\right),
\end{align*}
and $\Gamma(x)$ and $J_\alpha(x)$ denote the Gamma and Bessel functions, respectively.
The mode functions have been normalized in terms of the Klein--Gordon product:
\begin{align*}
  (f,g)_{\rm KG}:=i\int_{-\infty}^\infty dX [f^*\partial_Tg-(\partial_Tf^*)g],
\end{align*}
such that
\begin{align*}
  (f(p;T,X),f(p';T,X))_{\rm KG}&=-  (f^*(p;T,X),f^*(p';T,X))_{\rm KG}=\delta(p-p'),\\
  (f(p;T,X),f^*(p';T,X))_{\rm KG}&=0
\end{align*}
hold.
Since the mode function $f(p;T,X)$ approximates the plane wave 
in the Minkowski space--time
in the limit $T\to -\infty$:
\begin{align*}
  f(p;T,X)=(4\pi |p|)^{-1/2}e^{-i|p|T}e^{ipX}+O(e^{4\beta T}),
\end{align*}
we take $\{f(p;T,X)\}$ as the positive frequency solutions and expand the
quantum Klein--Gordon field in the form
\begin{align*}
  \phi=\int_{-\infty}^\infty dp~ [a(p)f(p;T,X)+a^*(p)f^*(p;T,X)].
\end{align*}
The Fock space $\mathscr{F}_\Omega$ is constructed from the vacuum state $|\Omega\rangle$ defined by the requirement:
\begin{align*}
  a(p)|\Omega\rangle=0,~~~\mbox{(for all $p\in\boldsymbol{R}$)},
\end{align*}
according to the standard procedure.

The computation of the quantum Hamiltonian operator for the Klein--Gordon field in this case 
leads to
\begin{align*}
  H(T)&=\int dp~\left[\sigma(|p|;T)a^*(p)a(p)
+\dfrac{\tau(|p|;T)}{2}a(p)a(-p)+\dfrac{\tau^*(|p|;T)}{2}a^*(p)a^*(-p)\right],
\end{align*}
where
\begin{align*}
  \sigma(|p|;T)&:=|\Gamma(1-i(2\beta)^{-1}|p|)|^2\\
&\times \biggl[
\dfrac{A^2m^2e^{4\beta T}}{8|p|}
\biggl|
J_{i(2\beta)^{-1}|p|-1}\left(\dfrac{Ame^{2\beta T}}{2\beta}\right)
-J_{i(2\beta)^{-1}|p|+1}\left(\dfrac{Ame^{2\beta T}}{2\beta}\right)
\biggr|^2\\
&+\dfrac{1}{2}\left(|p|+\dfrac{A^2m^2 e^{4\beta T}}{|p|}\right)
\biggl|
J_{i(2\beta)^{-1}|p|}\left(\dfrac{Ame^{2\beta T}}{2\beta}\right)
\biggr|^2
\biggr],\\
\tau(|p|;T)&:=\Gamma(1-i(2\beta)^{-1}|p|)^2
\left(\dfrac{Am}{4\beta}\right)^{i\beta^{-1}|p|}\\
&\times\biggl\{
\dfrac{A^2m^2 e^{4\beta T}}{8|p|}\biggl[
J_{-i(2\beta)^{-1}|p|-1}\left(\dfrac{Ame^{2\beta T}}{2\beta}\right)
-J_{-i(2\beta)^{-1}|p|+1}\left(\dfrac{Ame^{2\beta T}}{2\beta}\right)
\biggr]^2\\
&+\dfrac{1}{2}\left(|p|+\dfrac{A^2m^2e^{4\beta T}}{|p|}\right)
\left[J_{-i(2\beta)^{-1}|p|}\left(\dfrac{Ame^{2\beta T}}{2\beta}\right)\right]^2
\biggr\}.
\end{align*}

The Hamiltonian can be put into the diagonal form by the transformation~\cite{Par69,BCF78}
\begin{align}
  a(p;T)&=a(p)\cosh\theta (|p|;T)+a^*(-p)e^{i\gamma(|p|;T)}\sinh\theta (|p|;T),
\label{eq:bogoliubov}
\end{align}
where the real functions $\theta(|p|;T)$ and $\gamma(|p|;T)$ 
are determined by
\begin{align*}
  e^{2\theta(|p|;T)}&=\sqrt{\dfrac{\sigma(|p|;T)+|\tau(|p|;T)|}{\sigma(|p|;T)-|\tau(|p|;T)|}},\\
e^{-i\gamma(|p|;T)}&=\dfrac{\tau(|p|;T)}{|\tau(|p|;T)|}.
\end{align*}
Then, the Hamiltonian becomes 
\begin{align*}
  H=\int dp~ \omega(|p|;T)a^*(p;T)a(p;T),
\end{align*}
 where 
\begin{align*}
  \omega(|p|;T):=\sqrt{\sigma(|p|;T)^2-|\tau(|p|;T)|^2}.
\end{align*}

Taking the number operators $N(p;T):=a^*(p;T)a(p;T)$  as the
observables at the time $T$,  the evaluation of 
 $\langle \Omega |N(p;T)|\Omega\rangle$
gives a non--zero value.
Then, it is argued that it describes the creation of scalar particles
in the expanding universe,
which is the approach frequently taken in the context of the quantum
field theory in the FRW universe.
Here,  we seek for an appropriate formulation
of the quantum cosmology
assuming that we can observe only a 1--particle state of the universe
as a guiding principle.

Since the defining equation~(\ref{eq:bogoliubov}, section 3) of $a(p;T)$ has
the form of the Bogoliubov transformation, it can be also written as
\begin{align*}
  a(p;T)&=U(T)a(p)U^*(T),\\
U(T)&:=\exp\biggl\{
\dfrac{1}{2}\int dp~\theta(|p|;T)
[ e^{-i\gamma(|p|;T)}a(p)a(-p)-e^{i\gamma(|p|;T)}a^*(p)a^*(-p)]
\biggr\}.
\end{align*}
The unitary operator $U(T)$ is improper~\cite{Rom69} 
in the sense that it is not a unitary operator on the 
 Fock space $\mathscr{F}_\Omega$.
The Fock space $\mathscr{F}_T$ $(T\in\boldsymbol{R})$
can be built from the ``$T$--vacuum''
\begin{align*}
|\Omega;T\rangle:=U(T)|\Omega\rangle,  
\end{align*}
subject to $a(p;T)|\Omega;T\rangle=0$ $(p\in\boldsymbol{R}$),
by applying all polynomials of $a^*(p;T)$ according to the standard 
procedure for the Fock representation.
In this way, the 
$T$--vacuum $|\Omega;T\rangle$ is regarded as the instantaneous 
vacuum of the Klein--Gordon field at the finite time $T$,
and we have a continuum of mutually
improperly equivalent Fock spaces parametrized
by $T$.

Since the Hamiltonian $H(T)$ is a Hermitian
operator in $\mathscr{F}_T$, a finite particle state in $\mathscr{F}_T$ 
remains in a finite particle state under the action of 
the unitary operator $\exp(-iH(T)d T)$ in $\mathscr{F}_T$.
This is of course not the state in $\mathscr{F}_{T+d T}$.
Instead, it is physically interpreted as the infinite particle state in 
$\mathscr{F}_{T+d T}$.
This implies that the Klein--Gordon Hamiltonian $H(T)$ does not define 
a unitary time evolution in a separable Hilbert space.
In this sense, this theory deviates from the framework of 
the standard quantum theory.

This leads to a paradoxical conclusion that
an infinitely many universes are continuously created.
While in the context of the quantum field theory in the FRW background,
such divergence of the particle number might not be regarded as so problematic,
since its physical meaning is clearly interpretable.
In fact, this kind of divergence comes from the infinite spatial volume of
the background geometry, and the expectation value of
the  appropriate number density operator remains finite.
Nevertheless, in the present context of the quantum cosmology,
it is hard to see the correspondence to the
classical solution to the Einstein equation with such an interpretation,
because we always observe only a single--universe state in any case.  
Hence, an alternative framework for the quantum dynamics is required.

\begin{figure}[htbp]
\includegraphics[width=.8\linewidth]{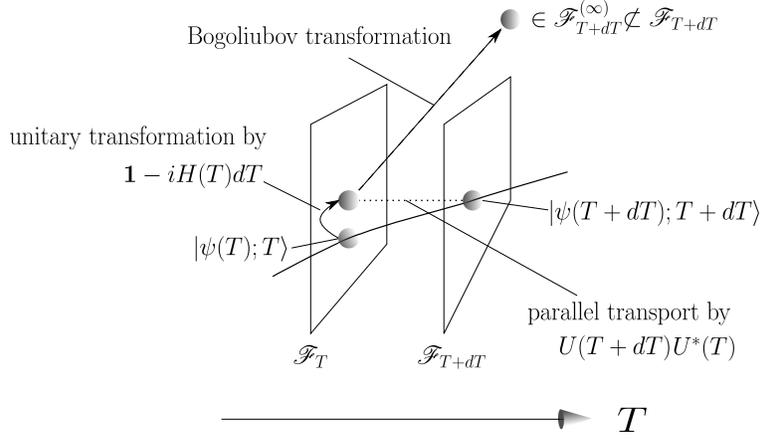}
\caption{Schematic picture of the vector bundle structure in the continuum
of Fock spaces. The dynamics of the quantum state is given by the
unitary transformation $\exp [-i H(T)dt]$ in $\mathscr{F}_T$ 
followed by the parallel transport defined by $U(T+dT)U^*(T)$.}
\label{fig:hilb}
\end{figure}

In order to provide a natural notion of the unitary time evolution,
we need to define a time derivative between states belonging to
mutually different Fock spaces. 
A natural way to realize this would be provided by 
considering the fibre bundle structure~\cite{Ste51} in the continuum of the Fock spaces
by 
\begin{align*}
\pi:\bigcup_{T\in\boldsymbol{R}} \mathscr{F}_T =:F\to \boldsymbol{R};|\psi;T\rangle \mapsto T,
\end{align*}
and the local trivialization of this vector bundle is supposed to be given by
\begin{align*}
\varphi: \boldsymbol{R}\times \mathscr{F}_\Omega\to F ;
(T,|\psi;\Omega\rangle)\mapsto |\psi;T\rangle:=U(T)|\psi;\Omega\rangle.
\end{align*}
In this setting, the dynamics of the quantum state can be described by a global section
\begin{align*}
  \psi: \boldsymbol{R}\to F; T\mapsto |\psi(T);T\rangle
\end{align*}
of $F$ (See Fig.~\ref{fig:hilb}).

In order to formulate the quantum dynamics, we have to define 
the time derivative of the motion $|\psi(T);T\rangle$,
where a problem is that $d/dT$ is not an anti--Hermitian operator
on the Fock space $\mathscr{F}_T$.
To make the time derivative of the quantum state a meaningful operation,
we need to introduce the notion of parallel transport of the quantum state
from $\mathscr{F}_T$ to $\mathscr{F}_{T+dT}$.
Then, our vector bundle structure provides a natural framework
to define the parallel transport in terms of the connection of 
the vector bundle.
The most natural choice of the parallel transport would be
given by 
\begin{align*}
u(T',T):=U(T')U^*(T):\mathscr{F}_T\to \mathscr{F}_{T'}.
\end{align*}
Now, we define the covariant time derivative of the
quantum state with respect to this parallel transport by
\begin{align*}
  D_T:=\dfrac{d}{dT}+U(T)(\partial_TU^*(T)),
\end{align*}
which is an anti--Hermitian operator on $\mathscr{F}_T$.
The second term appeared in $D_T$ is, in a sense, the gauge field,
which belongs to the Lie algebra of the structure group: the group
of all unitary operators on the Fock space $\mathscr{F}_\Omega$.
In terms of this covariant time derivative operator, we postulate
that the unitary time evolution of the quantum state is determined by
 the covariant Schr\"odinger equation\footnote{The general form of the
covariant Schr\"odinger equation is possibly given by
$(iD_T+A(T))|\psi(T);T\rangle=H(T)|\psi(T);T\rangle$ in terms of a
gauge field $A(T)$. Here we simply consider the case: $A(T)=0$, because 
an additional structure to determine $A(T)\ne 0$ is not equipped 
with the present settings.}
\begin{align*}
  iD_T|\psi(T);T\rangle=H(T) |\psi(T);T\rangle.
\end{align*}
In terms of the local coordinates in $F$,
this is equivalent to the Schr\"odinger equation 
\begin{align*}
  i\partial_T|\psi(T);\Omega\rangle
&=H_\Omega(T)|\psi(T);\Omega\rangle,\\
H_\Omega(T):&=\int dp~ \omega(|p|;T)a^*(p)a(p),
\end{align*}
in $\mathscr{F}_\Omega$.
Thus, in this formulation, 
everything is described in the language of a single separable
Hilbert space $\mathscr{F}_\Omega$.
 
Our quantum dynamics has an advantage that any 1--particle state
remains in a 1--particle state, which would be particularly 
suitable for the  quantum cosmology.
So, let us introduce the canonical observable set on the 1--particle
Fock spaces $\mathscr{F}^{(1)}_T$, along the line of the case of the flat universe.

Firstly, define the Hermitian operators
\begin{align*}
  P(T)&:=\int_{-\infty}^\infty dp~pa^*(p;T)a(p;T),\\
  Q(T)&:=i\int_{-\infty}^\infty dp~a^*(p;T)\partial_pa(p;T),
\end{align*}
on $\mathscr{F}^{(1)}_T$ are regarded as the momentum and 
position operators, respectively.
In fact, these realize the representation of the CCR--algebra
\begin{align*}
  [Q(T),P(T)]_-=i\boldsymbol{1} ,
\end{align*}
on the 1--particle Fock space $\mathscr{F}^{(1)}_T$.

The eigenvector of $P(T)$ is written as
\begin{align*}
  |p;T\rangle:=a^*(p;T)|\Omega;T\rangle,
\end{align*}
corresponding to the eigenvalue $p\in\boldsymbol{R}$,
and the eigenvector of $Q(T)$ is formally written as
\begin{align*}
  |X;T\rangle_Q:=(2\pi)^{-1/2}\int_{-\infty}^\infty dp~e^{-ipX}|p;T\rangle,
\end{align*}
which corresponds to the eigenvalue $X\in\boldsymbol{R}$.
Hence, $|X;T\rangle_Q$ can be regarded as the 1--particle state localized at the position
$X$ at time $T$.

Since $|X;T'\rangle_Q=u(T',T)|X;T\rangle_Q$ holds, 
the localized  1--particle state $|X;T'\rangle_Q\in \mathscr{F}^{(1)}_{T'}$ is 
the parallel transport of 
$|X;T\rangle_Q\in \mathscr{F}^{(1)}_{T}$ with the same position.
It gives
justification for our choice of the connection of the vector bundle.

From the completeness of the 1--particle states
\begin{align*}
  \int_{-\infty}^\infty dp~ |p;T\rangle\langle p;T|=\boldsymbol{1},
\end{align*}
in $\mathscr{F}^{(1)}_T$, we can expand the 1--particle state $|\psi(T);T\rangle$ as
\begin{align*}
  |\psi(T);T\rangle=\int_{-\infty}^\infty dp~ \psi(p,T)|p;T\rangle,
\end{align*}
and the coefficient $\psi(p,T)$ is the wave function in the momentum representation.
The Hamiltonian operator $H(T)$ restricted on the 1--particle Fock space $\mathscr{F}^{(1)}_T$
can be written as
\begin{align*}
  H^{(1)}(T)=\int_{-\infty}^\infty dp~ \omega( |p|,T) a^* (p;T) a(p;T).
\end{align*}
By calculating its matrix element, we obtain the Schr\"odinger equation
\begin{align}
\nonumber  i\partial_T\psi(p,T)&=\int_{-\infty}^\infty dp'~ \langle p;T|H^{(1)}(T)|p';T\rangle \psi(p',T)\\
&=\omega(|p|,T)\psi(p,T).
\end{align}
in the momentum representation on $\mathscr{F}^{(1)}_T$.
We show in section~\ref{sec:c-q} that this reproduces a correct classical dynamics.

\subsection{Case of elliptic universe $(K=1)$:}\label{subsec:ell}

The semi--Riemannian metric on $\mathscr{M}$ in this case is given by
[See Eq.~(\ref{eq:gc})]
\begin{align*}
  g_C=\dfrac{3 v^2}{(8\pi G)^2C^2} (12 a^2 da^2-a^4 dX^2).
\end{align*}
Introducing $T$ by  $a=a_0 e^{\beta T}$, this becomes
\begin{align*}
  g_C=A^2e^{4\beta T}(dT^2-dX^2),
\end{align*}
where
\begin{align*}
  \beta=(12)^{-1/2}, ~~A=\biggl|\dfrac{\sqrt{3}va_0^2}{8\pi G C}\biggr|.
\end{align*}
Since the classical Hamiltonian is given by
\begin{align*}
  H'_T=\lambda((g_C)^{mn}p_mp_n+1),
\end{align*}
a classical motion corresponds to a space--like geodesic in $\mathscr{M}$,
if we take $T$ as the time function.

As a quantum version of this classical system, we take the
tachyonic quantum Klein--Gordon field~\cite{Tan60} described by the action
\begin{align*}
S[\phi]=\dfrac{1}{2}\int dT dX [(\partial_T\phi)^2-(\partial_X\phi)^2+A^2m^2e^{4\beta T}\phi^2].
\end{align*}

The mode functions $\{f(p;T,X),f^*(p;T,X)\}$ for the Klein--Gordon equation
in this case can be written as
 \begin{align*}
  f(p;T,X)&:=(4\pi |p|)^{-1/2}F(|p|;T)e^{ip X},\\
F(|p|;T)&:=\Gamma(1-i(2\beta)^{-1}|p|)
\left(\dfrac{Am}{4\beta}\right)^{i(2\beta)^{-1}|p|}
I_{-i(2\beta)^{-1}|p|}\left(\dfrac{Am e^{2\beta T}}{2\beta}\right),
\end{align*}
where $I_\alpha(x)$ denotes the modified Bessel function of the first kind.
Since the mode function $f(p;T,X)$ becomes the plane wave 
\begin{align*}
  f(p;T,X)=
(4\pi |p|)^{-1/2}e^{-i|p|T}e^{ipX}+O(e^{4\beta T}),
\end{align*}
as $T\to -\infty$, 
we regard $\{f(p;T,X);p\in \boldsymbol{R}\}$ 
as the positive frequency solutions.

The subsequent procedure to obtain the first--quantized theory
is parallel to the $K=-1$ case. 
However, we need to specify the functional space of the wave functions 
to define the unitary dynamics of the 1--particle quantum state of the
tachyonic field.

Firstly,  quantize the Klein--Gordon field 
\begin{align*}
  \phi(T,X)=\int dp~ [a (p) f(p;T,X)+a^*(p) f^*(p;T,X)],
\end{align*}
with the operators $\{a(p),a^*(p);p\in \boldsymbol{R}\}$. 
In the case of the tachyonic scalar field in the Minkowski background,
it is known that the CCRs for the scalar field is not compatible with
the harmonic--oscillator commutation relations for the creation and
annihilation operators~\cite{Fei67}. 
It however turns out that we don't encounter this kind of difficulties 
in the present background geometry.
Here,
 the CCRs for $\{\phi(T,X),\partial_T\phi(T,X);X\in\boldsymbol{R}\}$
are equivalent to
the
harmonic--oscillator commutation relations
\begin{align*}
  [a(p),a^*(p')]_-&=\delta(p-p')\boldsymbol{1},~~
  [a(p),a(p')]_-=0,~~
  [a^*(p),a^*(p')]_-=0,
\end{align*}
for the operators $\{a(p),a^*(p);p\in\boldsymbol{R}\}$.
The vacuum state is defined by $a(p)|\Omega\rangle=0$ $(p\in\boldsymbol{R})$
and we obtain the Fock representation on $\mathscr{F}_\Omega$ 
in the standard procedure.

Then, the Hamiltonian operator becomes
\begin{align*}
  H(T)&=\int dp~ h(p;T),\\
\end{align*}
where
\begin{align*}
h(p;T)&:=\sigma(|p|;T)a^*(p)a(p)
+\dfrac{\tau(|p|;T)}{2}a(p)a(-p)+\dfrac{\tau^*(|p|;T)}{2}a^*(p)a^*(-p),\\
  \sigma(|p|;T)&:=|\Gamma(1-i(2\beta)^{-1}|p|)|^2\\
&\times \biggl[
\dfrac{A^2m^2e^{4\beta T}}{8|p|}
\biggl|
I_{i(2\beta)^{-1}|p|-1}\left(\dfrac{Ame^{2\beta T}}{2\beta}\right)
+I_{i(2\beta)^{-1}|p|+1}\left(\dfrac{Ame^{2\beta T}}{2\beta}\right)
\biggr|^2\\
&+\dfrac{1}{2}\left(|p|-\dfrac{A^2m^2 e^{4\beta T}}{|p|}\right)
\biggl|
I_{i(2\beta)^{-1}|p|}\left(\dfrac{Ame^{2\beta T}}{2\beta}\right)
\biggr|^2
\biggr],\\
\tau(|p|;T)&:=\Gamma(1-i(2\beta)^{-1}|p|)^2
\left(\dfrac{Am}{4\beta}\right)^{i\beta^{-1}|p|}\\
&\times\biggl\{
\dfrac{A^2m^2 e^{4\beta T}}{8|p|}\biggl[
I_{-i(2\beta)^{-1}|p|-1}\left(\dfrac{Ame^{2\beta T}}{2\beta}\right)
+I_{-i(2\beta)^{-1}|p|+1}\left(\dfrac{Ame^{2\beta T}}{2\beta}\right)
\biggr]^2\\
&+\dfrac{1}{2}\left(|p|-\dfrac{A^2m^2e^{4\beta T}}{|p|}\right)
\left[I_{-i(2\beta)^{-1}|p|}\left(\dfrac{Ame^{2\beta T}}{2\beta}\right)\right]^2
\biggr\}.
\end{align*}
The Hamiltonian at the time $T$ can be made in a diagonalized form
in the integration range
$|p|> p_0(T)$,
 where the function $p_0(T)$ is 
the smallest positive solution of the equation
\begin{align*}
  \sigma(p_0(T);T)=|\tau(p_0(T);T)|,
\end{align*}
that is given by
\begin{align*}
  p_0(T)=Am e^{2\beta T}.
\end{align*}
Keeping this in mind, we separate the Hamiltonian
into two parts as
\begin{align*}
  H(T)=\left(\int_{|p| > \xi} +\int_{|p|\le \xi}\right)dp~h(p;T),
\end{align*}
in terms of an arbitrarily fixed positive parameter $\xi$.
The first part of the Hamiltonian 
can be diagonalized via the Bogoliubov transformation
\begin{align*}
  a(p;T)&=U(T)a(p)U^*(T),\\
U(T)&:=\exp\biggl\{
\dfrac{1}{2}\int_{|p|> \xi} dp~\theta(|p|;T)
[ e^{-i\gamma(|p|;T)}a(p)a(-p)-e^{i\gamma(|p|;T)}a^*(p)a^*(-p)]
\biggr\},\\
  e^{2\theta(|p|;T)}&:=\sqrt{\dfrac{\sigma(|p|;T)+|\tau(|p|;T)|}{\sigma(|p|;T)-|\tau(|p|;T)|}},\\
e^{-i\gamma(|p|;T)}&:=\dfrac{\tau(|p|;T)}{|\tau(|p|;T)|}.
\end{align*}
The above expressions are valid
for the range $T\in (-\infty, T_1)$,
where $T_1$ is determined by
$\xi=p_0(T_1)$, i.e. $T_1=(2\beta)^{-1}\log (\xi/Am)$.

In this way, we have a partly diagonalized form of the Hamiltonian
\begin{align*}
  H(T)&=\int_{|p|>\xi} dp~\omega(|p|;T)a^*(p;T)a(p;T)+\int_{|p|\le \xi}dp~h(p;T),\\
\omega(|p|;T)&:=\sqrt{\sigma(|p|;T)^2-|\tau(|p|;T)|^2}.
\end{align*}
where $\omega(|p|;T)$ is a real function, hence the first diagonal part is
a Hermitian operator on $\mathscr{F}_T$ for $T< T_1$.

Accordingly, as shown in the following, 
we can at best obtain the unitary theory of the 1--particle states
within the time duration $T\in(-\infty, T_1)$ for $K=1$ universe. 
This would correspond to the fact that each classical solution describes
a recollapsing universe, and hence the scale factor $a=a_0e^{\beta T}$
always has an upper bound.

To describe this system, we regard the Fock space $\mathscr{F}_\Omega$ 
as the tensor product space $\mathscr{A}\otimes \mathscr{B}$
of the subsystems $\mathscr{A}$ and $\mathscr{B}$, where
$\mathscr{A}$ (resp. $\mathscr{B}$) is the Fock space with respect to the
creation and annihilation operators $\{a(p),a^*(p);|p|> \xi\}$
(resp. $\{a(q),a^*(q);|q|\le \xi\}$).
The $T$--vacuum is defined by
\begin{align*}
a(p;T)|\Omega;T\rangle&=0,~~\mbox{for $|p|>\xi$}\\
a(q)|\Omega;T\rangle&=0,~~\mbox{for $|q|\le\xi$}
\end{align*}
and the Fock space $\mathscr{F}_T$ is constructed with respect to this vacuum
by applying polynomials of $\{a^*(p;T),a^*(q);|p|>\xi,|q|\le\xi\}$.
The Fock space $\mathscr{F}_T$ can be regarded as the tensor product
of a pair of Fock spaces $\mathscr{A}_T$ and $\mathscr{B}$,
$\mathscr{A}_T$ (resp. $\mathscr{B}$) being the Fock representation space 
of the algebra generated by $\{a(p;T), a^*(p;T);|p|> \xi\}$
(resp. $\{a(q), a^*(q);|q|\le \xi\}$).
In other words, $\mathscr{F}_T$ is spanned by the vectors in the form
\begin{align*}
  |n_{p_1},n_{p_2},\cdots,n_{p_r};T\rangle\otimes
  |n_{q_1},n_{q_2},\cdots,n_{q_s}\rangle,
\end{align*}
where 
$|n_{p_1},n_{p_2},\cdots,n_{p_r};T\rangle\in \mathscr{A}_T$
(resp. $|n_{q_1},n_{q_2},\cdots,n_{q_s}\rangle\in \mathscr{B}$)
denotes the simultaneous eigenvector of the number operators
$a^*(p;T)a(p;T)$ $(|p|> \xi)$ [resp. $a^*(q)a(q)$ $(|q|\le \xi)$].
Accordingly, the Hamiltonian $H(T)$ is 
regarded as the sum
\begin{align*}
  H(T)&=H_\xi(T)\otimes \boldsymbol{1}
+ \boldsymbol{1}\otimes\int _{|p|\le\xi}dp~h(p;T),\\
  H_\xi(T)&:=\int_{|p|> \xi}dp~\omega(|p|;T)a^*(p;T)a(p;T),
\end{align*}
of the operator on $\mathscr{A}_T$ and 
the operator on $\mathscr{B}$.
Hence, for a vector state $|\psi\rangle \otimes |\psi'\rangle\in
\mathscr{A}_T\otimes \mathscr{B}$, each vector
 $|\psi\rangle$
(resp. $|\psi'\rangle$) independently undergoes the unitary time evolution
in the inseparable Fock space $F_\xi:=\bigcup_{T< T_1}\mathscr{A}_T$ (resp. $\mathscr{B}$).
Hence, we can consistently assume that our observable set consists only of the
Hermitian operators of the form $O(T)\otimes \boldsymbol{1}$,
In other words,  we can restrict ourselves to the 
dynamics of the subsystem $F_\xi$.

As in the case of $K=-1$, the Hilbert spaces $\mathscr{A}_T$ 
and $\mathscr{A}_{T'}$ 
are improperly unitary equivalent, when $T\ne T'$.
In a similar procedure to the $K=-1$ case,
we introduce the vector bundle structure by
the projection 
\begin{align*}
\pi:  \bigcup_{T< T_1}\mathscr{A}_T=:F_\xi\to (-\infty,T_1);
|\psi;T\rangle\mapsto T,
\end{align*}
and the local trivialization
\begin{align*}
  \varphi: (-\infty,T_1)\times \mathscr{A}_\Omega\to F;
(T,|\psi;\Omega\rangle)\mapsto 
|\psi;T\rangle:=U(T)|\psi;\Omega\rangle.
\end{align*}

Our Hilbert space describing the 1--particle state of the universe
at the time $T$
is assumed to be the subspace $\mathscr{A}^{(1)}_T\otimes \mathscr{B}$,
where $\mathscr{A}^{(1)}_T$ denotes the space of 1--particle states in
 $\mathscr{A}_T$. Each such state is projected onto the space of 
quantum state on $\mathscr{A}_T$ via the partial trace over
the subsystem $\mathscr{B}$,
 which may produce a mixed state on $\mathscr{A}_T$. 
Since the structure of the Hamiltonian ensures that the operation of the partial trace
and the unitary time evolution by the Hamiltonian 
commute,
the quantum dynamics is reduced to that of a separable state,
which can be represented as a vector state consisting of a single term
 $|\psi;T\rangle\otimes |\psi'\rangle$.
Such a separable state simply corresponds to the vector state $|\psi;T\rangle\in \mathscr{A}_T$,
and its time evolution is assumed to described by the Schr\"odinger equation
\begin{align*}
iD_T|\psi(T);T\rangle&=  H_\xi(T)|\psi(T);T\rangle,\\
D_T&:={d\over dT}+U(T)(\partial_T U(T)^*),
\end{align*}
for $T\in (-\infty,T_1)$.

The canonical observable set $\{Q(T),P(T)\}$ on $\mathscr{A}^{(1)}_T$
are also constructed in a similar manner.
Firstly, we define the momentum operator by
\begin{align*}
  P(T):=\int_{|p|>\xi} dp~p~ a^*(p;T) a(p;T),
\end{align*}
which is a Hermitian operator on $\mathscr{A}^{(1)}_T$.
 The eigenstate of $P(T)$ is given by
 \begin{align*}
   |p;T\rangle&:=a^*(p;T)|\Omega_{\mathscr{A}};T\rangle,
\end{align*}
which satisfies
\begin{align*}
P(T)|p;T\rangle&=p|p;T\rangle,
 \end{align*}
where $|\Omega_{\mathscr{A}};T\rangle$ denotes the vacuum state 
in $\mathscr{A}_T$.
Next, the position operator is defined by
\begin{align*}
  Q(T):=i\int_{|p|>\xi}dp~a^*(p;T)\partial_p a(p;T), 
\end{align*}
which is a Hermitian operator on $\mathscr{A}^{(1)}_T$.
These constitute the CCR--algebra
\begin{align*}
  [Q(T),P(T)]_-=i\boldsymbol{1},
\end{align*}
where we abbreviate the identity operator on $\mathscr{A}^{(1)}_T$ as
\begin{align*}
  \boldsymbol{1}:=\int_{|p|> \xi}dp~a^*(p;T)a(p;T).
\end{align*}
The eigenvector of $Q(T)$ can be written as
\begin{align*}
|X;T\rangle_Q&:=(2\pi)^{-1/2}\int_{|p|>\xi}dp~e^{-ipX}|p;T\rangle,
\end{align*}
for $X\in\boldsymbol{R}$, and it satisfies
\begin{align*}
    Q(T)|X;T\rangle_Q&=
X|X;T\rangle_Q.
\end{align*}
Unlike the $K=-1$ case, $\{|X;T\rangle_Q;X\in\boldsymbol{R}\}$
does not satisfy the orthogonality condition, i.e. 
it holds
\begin{align*}
  {}_Q\langle X;T|X';T\rangle_Q\ne \delta(X-X'),
\end{align*}
in the present case. 
However,
since for the operator $F_X(T):=|X;T\rangle_Q  {}_Q\langle X;T|$, it
hold
\begin{align*}
\langle \psi;T|F_X(T)|\psi;T\rangle&>0,~~~\mbox{for all $|\psi;T\rangle\in \mathscr{A}^{(1)}_T$}\\
  \int_{-\infty}^\infty dX F_X(T)&=\boldsymbol{1},
\end{align*}
the measurement of the position operator 
\begin{align*}
  Q(T)=\int dX~ X F_X(T)
\end{align*}
can be regarded as a POVM measurement, though not a von Neumann measurement.

In the momentum representation on the 
1--particle Fock space $\mathscr{A}^{(1)}_T$,
the 1--particle state is described by
the wave function $\psi(p,T):=\langle p;T|\psi(T);T\rangle$,
and it is subject to the Schr\"odinger equation
\begin{align*}
  i\partial_T\psi(p,T)&=\omega(|p|,T)\psi(p,T).
\end{align*}
We see in the next section that the quantum system described here reproduces 
the correct classical theory.

\section{Classical--quantum correspondence
in the first--quantized theory}\label{sec:c-q}

Here, we show that the first--quantized theory obtained in the
previous section correctly reproduces the classical Einstein gravity.

In the previous section, we have obtained a quantum theory
of the expanding universe described by the wave function 
$\psi(p,T)$ of the universe subject to the Schr\"odinger equation
The expression of $\omega(|p|;T)$ includes the Bessel functions when $K=\pm 1$, which makes it difficult to solve the Schr\"odinger equation strictly. 
Nevertheless we can solve it approximately in terms of asymptotic forms of the Bessel functions. 
The Bessel function and the modified Bessel function in the mode functions 
are expanded around their argument $Am e^{2\beta T}=0$ as
\begin{align*}
&J_{-i(2\beta)^{-1}|p|}\left({A me^{2\beta T}\over 2\beta}\right)=\sum_{n=0}^\infty{(-1)^n\over n!\Gamma(n-i(2\beta)^{-1}|p|+1)}\left({A me^{2\beta T}\over 4\beta }\right)^{2n-i(2\beta)^{-1}|p|}, \\
&I_{-i(2\beta)^{-1}|p|}\left({A me^{2\beta T}\over 2\beta}\right)=\sum_{n=0}^\infty{1\over n!\Gamma(n-i(2\beta)^{-1}|p|+1)}\left({A me^{2\beta T}\over 4\beta }\right)^{2n-i(2\beta)^{-1}|p|}.
\end{align*}
From these,  it can be seen that 
the series expansion of $\omega(|p|;T)$ 
becomes
\begin{align*}
\omega(|p|;T)&=\sqrt{p^2-KA^2m^2 e^{4\beta T}}
+K{3(16+19p^2-45p^4+18p^6)\over 32|p|^5(1+3p^2)^3(4+3p^2)}
A^6m^6e^{12\beta T}\\
& +O(e^{16\beta T}).
\end{align*}
For $K=-1$, this expression is valid for $p\ne 0$,
and for $K=1$, it is valid for $|p|>A m e^{2\beta T}$. 
Thus, our Hamiltonian operator $\omega(|p|,T)$ well approximates
the naive Hamiltonian
\begin{align*}
  H_{\rm c}:=\sqrt{p^2-KA^2m^2 e^{4\beta T}},
\end{align*}
for $e^{2\beta T}\ll (Am)^{-1}$.
The classical counterpart of this naive Hamiltonian exactly gives 
 the equation of motion for the classical trajectory.

This correspondence between our Hamiltonian $\omega(|p|,T)$ and the naive Hamiltonian $H_{\rm c}$
holds wider range of $T$.
To see this, we next consider the late time behavior of $\omega(|p|;T)$.
For $e^{2\beta T}\gg (Am)^{-1}$ and $p\ne 0$, the Bessel function in the mode function 
behaves as
\begin{align*}
J_{-i(2\beta)^{-1}|p|}\left({A me^{2\beta T}\over 2\beta}\right)&=\sqrt{4\beta \over \pi A m}e^{-\beta T}
\Bigl[
\Bigl\{1-{3(1+12p^2)(3+4p^2)\over 128A^2m^2}e^{-4\beta T}\Bigl\}\\
& \times\cos\left(
{A m\over 2\beta }e^{2\beta T}-{\pi\over 4}+{i\pi|p|\over 4\beta }
\right)\\
& 
+{\beta(1+12p^2)\over 4A m}e^{-2\beta T}\sin\left(
{A m\over 2\beta }e^{2\beta T}-{\pi\over 4}+{i\pi|p|\over 4\beta }
\right)
\Bigl]+O(e^{-7\beta T}),
\end{align*}
which yields the behavior of $\omega(|p|;T)$ for $K=-1$, 
\begin{align*}
\omega(|p|;T)&=Ame^{2\beta T}+{1\over 2}{e^{-2\beta T}\over Am \sinh^2 (\pi|p|/(2\beta))}\Bigl[{(1-12p^2)\over 12}\\
&+{5+48p^4\over 32}\Bigl\{\cos\left({A m\over \beta}e^{2\beta T}-{\pi\over 2} \right)\cosh{\pi |p|\over 2\beta}-1\Bigl\}\\
&+2p^2\cos\left({A m\over \beta}e^{2\beta T}-{\pi\over 2} \right)\cosh{\pi |p|\over 2\beta}\Bigl]
+O(e^{-4\beta T}). 
\end{align*}
On the other hand,  the naive Hamiltonian $H_{\rm c}$ for $K=-1$ is 
expanded for $e^{2\beta T}\gg (Am)^{-1}$ as
\begin{align*}
H_{\rm c}&=\sqrt{p^2+A^2m^2e^{4\beta T}}\\
&=Ame^{2\beta T}+{p^2\over 2A^2m^2}e^{-2\beta T}+O(e^{-6\beta T}).
\end{align*}
This shows $\omega(|p|;T)$ approximates $H_{\rm c}$ in the limit $T\to\infty$,
\begin{align*}
\dfrac{\omega(|p|;T)  -H_{\rm c}}{H_{\rm c}}=O\left(e^{-2\beta T}\right),
\end{align*}
for $p\ne 0$.

For the intermediate range of $T$, i.e. for $Am e^{2\beta T}=O(1)$, 
we can see the good numerical coincidence between $\omega(|p|,T)$ 
and $H_{\rm c}$. A similar argument also holds for the closeness between
$\partial_p\omega(|p|,T)$
and $\partial_pH_{\rm c}$.
Thus, we conclude that 
\begin{align*}
  \omega(|p|;T)\approx \sqrt{p^2-KA^2m^2e^{4\beta T}},\\
\partial_p  \omega(|p|;T)\approx \dfrac{p}{\sqrt{p^2-KA^2m^2e^{4\beta T}}},
\end{align*}
holds for $K=0,\pm 1$, and for all ranges of $T$ in consideration.

Next, we show that the approximate Schr\"odinger equation
\begin{align}
\label{eq:schK}
  i\partial_T\psi(p;T)=\sqrt{p^2-KA^2m^2e^{4\beta T}}\psi(p;T)
\end{align}
applied to the wave packet state,
reproduces the  geodesic motion in the mini--superspace,
which corresponds to the classical solution to the Einstein equation.

Firstly, we note that the Ehlenfest--type theorem can be applied in the present case.
The expectation values of the momentum  and position operators for the normalized state $\psi(p,T)$
are given by
\begin{align*}
  \langle P\rangle&=\int dp~p\psi^* (p,T)\psi(p,T),\\
  \langle Q\rangle&=i\int dp~\psi^* (p,T)\partial_p\psi(p,T),
\end{align*}
where the range of integration is $p\in \boldsymbol{R}$ for $K=0,-1$, and $|p|>\xi$ for $K=1$.
The time derivative of these expectation values are readily obtained as
\begin{align*}
  \dfrac{d}{dT}\langle P\rangle&=0,\\
  \dfrac{d}{dT}\langle Q\rangle&=\left\langle \dfrac{\partial H_{\rm c}}{\partial p}\right\rangle
:=\int dp~\dfrac{\partial H_{\rm c}}{\partial p}\psi^* (p,T)\psi(p,T).
\end{align*}
This shows formal, though not strict, correspondence between
quantum and classical dynamics.  

Next, we consider the dynamics of the wave packet state.
The solution of Eq.~(\ref{eq:schK})  can be written as
\begin{align*}
\psi_k(p,T)
&=\exp\left(-i\int^Tds\sqrt{k^2-KA^2m^2e^{2\beta s}}\right)
\delta(p-k),
\end{align*}
where the solutions are labeled by 
the simultaneous eigenvalue $k$ of the momentum operator
subject to
$k\in \boldsymbol{R}$ for $K=0,-1$,
and  $|k|>\xi$ for $K=1$.
The general solution is written as
\begin{align*}
  \psi(p,T)=\int dk~ c(k) \psi_k(p,T),
\end{align*}
in terms of a coefficient $c(k)$ subject to the normalization condition
\begin{align*}
  \int dk~|c(k)|^2=1.
\end{align*}
Now, we consider the position representation of the wave function.
This is given by
\begin{align*}
  \widetilde\psi(X,T)&:={}_Q\langle X;T|\psi(T);T\rangle\\
&=(2\pi)^{-1/2}\int dp~ e^{ipX}\psi(p,T),
\end{align*}
so that it is generally written as
\begin{align}
\label{eq:wavepacket}
&\widetilde  \psi(X,T)
=(2\pi)^{-1/2}\int dk~ c(k) \exp\left(-i\int^Tds\sqrt{k^2-KA^2m^2e^{4\beta s}}\right)e^{ikX}.
\end{align}
The Born rule here could be stated as that 
$|\widetilde \psi(X,T)|^2$ gives the probability distribution that 
a measurement of the position operator
 at  time $T$ yields the value $X\in\boldsymbol{R}$. 

A wave packet state corresponding to a geodesic particle in mini--superspace
is obtained if we take $c(k)$ as a bell curve, say a Gaussian--like function,
 centered at $k=k_0\ne 0$, 
with the appropriately broad width $\sigma^{-1}$.
Then, $\widetilde\psi(X,T)$ describes a wave packet with the width $\sigma$.

From Eq.~(\ref{eq:wavepacket}), we can read off the
time dependence of the dispersion relation
\begin{align*}
\omega'(k,T)=\sqrt{k^2-KA^2m^2 e^{4\beta T}},
\end{align*}
between the wave--number $k$ and the instantaneous angular 
frequency $\omega'$.
Then, we readily find
that the group velocity $v_{\rm q}(k_0,T)$ of a
wave packet with the central wave--number $ k_0\ne 0$ is given by
\begin{align*}
  v_{\rm q}(k_0,T)&=\dfrac{\partial\omega'(k,T)}{\partial k}\biggl|_{k=k_0}\\
&=\dfrac{k_0}{\sqrt{k_0^2-KA^2m^2 e^{4\beta T}}}.
\end{align*}

The corresponding quantity can be obtained from the classical theory.
From the Hamiltonian constraint
\begin{align*}
 p_T^2-p_X^2 +KA^2 e^{4\beta T}\approx 0,
\end{align*}
we get the coordinate velocity of the geodesic particle as
\begin{align*}
  v_{\rm c}&=\dfrac{dX/ds}{dT/ds}=-\dfrac{p_X}{p_T}\approx
\dfrac{p_X}{\sqrt{p_X^2 -KA^2 e^{4\beta T}}}.
\end{align*}
This shows good agreement between the classical and quantum predictions
if we identify the dimensionless momentum $mp_X$ of the geodesic motion with 
the central wave--number $k_0$ of the wave packet state.

\section{Concluding remarks}\label{sec:final}
The quantum mechanical model of the cosmology 
studied here
is conceptually similar to the third--quantization model,
but technically different in that
 we respect the specific semi--Riemannian structure $(\mathscr{M},g_C)$ on 
the mini--superspace $\mathscr{M}$,
where the classical solution to the Einstein equation 
is given by the geodesic motion.
In the case of the Einstein gravity coupled with a
massless scalar field
in the FRW background,
the mini--superspace as a semi--Riemannian manifold
becomes a two--dimensional expanding universe.
Then, the classical solution to the Einstein equation is given by
a null (resp. time--like, space--like) geodesic in
the mini--superspace
for the flat (resp. hyperbolic, elliptic) 
FRW space--time.
Hence, we consider the quantum Klein--Gordon field
in the mini--superspace,
which becomes massless (resp. massive, tachyonic) for
the flat (resp. hyperbolic, elliptic) universe,
as a quantized model of the geodesic motion. 

In the case of the massive or tachyonic field,
we are faced with a known problem 
in the quantum field theory in a dynamical space--time
that 
an inseparable Hilbert space is required to describe
the quantum states of the field, where
the Klein--Gordon Hamiltonian at different time belongs to
a different Fock space. This continuum of the Fock spaces $\mathscr{F}_T$
parametrized by $T\in \boldsymbol{R}$ constitutes the inseparable Hilbert space,
which is too big.
By introducing the vector bundle structure 
in the continuum of the Fock space and the connection of the vector bundle,
we define the covariant time derivative for the quantum states,
in which the Klein--Gordon Hamiltonian can be regarded as a Hermitian operator
in a separable Hilbert space.
Furthermore, this framework gives a unitary time evolution of  1--particle states,
which would be preferable in the context of the quantum cosmology.
However, in the case of elliptic universe, the momentum space for
the Hilbert space of 1--particle states has to be restricted.
Then, we obtain the unitary theory for a restricted range of the time parameter
$T\in (- \infty,T_1)$.
 This time parameter corresponds to the scale factor of the
elliptic universe, which represents the recollapsing universe in the classical theory, 
so that this limitation of the  time parameter would correspond to
the existence of the upper bound for the scale factor.
It would not imply that the closed universe does not recollapse,
but rather it should be taken as
an indication of the limit of applicability of our model.
This might be a common issue of the quantum cosmological model based on the minisuperspace,
where the time function is given by the scale factor of the universe.

We construct the Hilbert space for the 1--particle states
and  the canonical observable set constituting the CCR--algebra
on the space of 1--particle states.
Accordingly, we obtain the Schr\"odinger equation for the wave function,
which is the Schr\"odinger representation of a 1--particle state.
The present quantization scheme
is free from the operator ordering ambiguities
and the problem of time unlike the Wheeler--DeWitt quantization scheme.
We find that the Hamiltonian is close to the naive Hamiltonian predicted from the Klein--Gordon
equation but with a small correction in the non--flat background cases.
We see that this ensures that our quantum theory correctly reproduces the Einstein equation.

A possible advantage of the present formalism 
over other third--quantized models is that it is applicable 
to the fermionic field
by considering the quantum Dirac field on $(\mathscr{M},g_C)$. 
Another good point is that we can treat the dynamics of the single--universe state, while in typical third--quantization models~\cite{McG88,Ban88,GS89,Wal93,HW95,Per12},
we always suffer from creation of infinite number of 
universes~\cite{Rub88,HM89,Vil94}.
However, we have to introduce a mass scale of the quantum field, 
which is a disadvantage of the theory. In fact, we have left an unknown dimensionless parameter
$Am$ in our formulation, which should be determined 
by a more fundamental theory.

\section*{Acknowledgments}
We would like to thank Takahiro Okamoto for useful discussions.

\end{document}